\documentclass{article}

\usepackage[position, final]{neurips_2025}

\usepackage[utf8]{inputenc} 
\usepackage[T1]{fontenc}    
\usepackage{hyperref}       
\usepackage{url}            
\usepackage{booktabs}       
\usepackage{amsfonts}       
\usepackage{nicefrac}       
\usepackage{microtype}      
\usepackage{xcolor}         

\usepackage{microtype}
\usepackage{graphicx}
\usepackage{subfigure}
\usepackage{booktabs} 

\usepackage{hyperref}
\usepackage{tabularx}

\usepackage{amsmath}
\usepackage{amssymb}
\usepackage{mathtools}
\usepackage{amsthm}
\usepackage{xurl}

\usepackage[capitalize,noabbrev]{cleveref}

\theoremstyle{plain}

\theoremstyle{definition}

\theoremstyle{remark}

\usepackage[textsize=tiny]{todonotes}

\title{Position Paper: If Innovation in AI Systematically Violates Fundamental Rights, Is It Innovation at All?}

\author{%
  Josu A. Eguiluz \\
  Adevinta ServicesCo S.L.U.\\Pompeu Fabra University (UPF) \\
  Ciutat de Granada, 150, Barcelona \\
  \texttt{josu.eguiluz@adevinta.com} \\
  \And
  Axel Brando \\
  Barcelona Supercomputing Center (BSC-CNS) \\
  Plaça d'Eusebi Güell, 1-3, Barcelona \\
  \texttt{axel.brando@bsc.es} \\
  \And
  Migle Laukyte \\
  Pompeu Fabra University (UPF) \\
  Ramon Trias Fargas, 25, Barcelona \\
  \texttt{migle.laukyte@upf.edu} \\
  \And
  Marc Serra-Vidal \\
  Kleinanzeigen.de GmbH (Adevinta) \\
  Ciutat de Granada, 150, Barcelona \\
  \texttt{marc.serrav@adevinta.com} \\
}

\begin{document}

\maketitle

\begin{abstract}
  
Artificial intelligence (AI) now permeates critical infrastructures and decision-making systems where failures produce social, economic, and democratic harm. This position paper challenges the entrenched belief that regulation and innovation are opposites. As evidenced by analogies from aviation, pharmaceuticals, and welfare systems and recent cases of synthetic misinformation, bias and unaccountable decision-making, the absence of well-designed regulation has already created immeasurable damage. Regulation, when thoughtful and adaptive, is not a brake on innovation—it is its foundation. The present position paper examines the EU AI Act as a model of risk-based, responsibility-driven regulation that addresses the Collingridge Dilemma: acting early enough to prevent harm, yet flexibly enough to sustain innovation. Its adaptive mechanisms—regulatory sandboxes, small and medium enterprises (SMEs) support, real-world testing, fundamental rights impact assessment (FRIA)—demonstrate how regulation can accelerate responsibly, rather than delay, technological progress. The position paper summarises how governance tools transform perceived burdens into tangible advantages: legal certainty, consumer trust, and ethical competitiveness. Ultimately, the paper reframes progress: innovation and regulation advance together. By embedding transparency, impact assessments, accountability, and AI literacy into design and deployment, the EU framework defines what responsible innovation truly means—technological ambition disciplined by democratic values and fundamental rights.
\end{abstract}

\section{Introduction}
\label{submission}

In recent years, Artificial Intelligence (AI) has gained remarkable traction in fields ranging from specialized
research domains to becoming indispensable drivers of economic growth, social transformation, and scientific discovery. Advocates of deregulation argue that reducing legal constraints fuels rapid innovation {\cite{APNews2025, BoschCEO2025, MITSloan2023, McAfee2021, Loten2021}}. Yet this position overlooks a crucial insight from Collingridge's Dilemma \cite{collingridge1980}: while the need for governance remains less obvious in the early stages of a technology, once its societal consequences become apparent, the technology is often too deeply embedded to be meaningfully altered by regulation and other normative instruments. \textbf{In this position paper, we argue that regulation is not only necessary but also a fundamental enabler of innovation, particularly in critical AI-based systems.} 

Drawing on historical precedents—ranging from pharmaceuticals to welfare systems or aviation—this position paper demonstrates how robust regulation has frequently fostered, rather than hindered, significant advancements. AI is not different: left unregulated, it can entrench harmful risks such as bias and discrimination, and unaccountable decision-making can have high-stake consequences. By examining the proposed EU AI Act as both an illustrative example and a case study, we explore how a risk-based approach to governance \cite{degregorio2022}, combined with mechanisms like regulatory sandboxes, can unlock technical breakthroughs while safeguarding fundamental rights. In fact, neglecting sustainable governance today is not just a regulatory risk—it might be an existential threat to the future of humankind \footnote{\textit{``Mitigating the risk of extinction from AI should be a global priority alongside other societal-scale risks such as pandemics and nuclear war.''} \cite{safeai2023} The 2025 UN ``Red Lines on AI'' initiative \cite{RedLines2025} echoed this call by urging global ethical boundaries to prevent catastrophic risks from high-impact AI systems.}. Ultimately, we ask whether innovations that systematically violate human dignity and perpetuate inequities can truly be labeled “innovative,” and we urge stakeholders to view legislation not as a barrier, but as a cornerstone for AI’s long-term viability.

\paragraph{Paper Structure} This position paper is organized into seven sections. Section 2 explains why the tension between regulation and innovation constitutes a false dichotomy, drawing on historical precedents and on the Collingridge Dilemma to show why late-stage regulation often arrives too late to prevent systemic harm. Section 3 examines the concrete risks of deregulated AI, from large-scale disinformation to bias and unaccountable decision-making in high-stakes contexts. Section 4 presents the EU AI Act as a model for risk-based, innovation-enabling governance, detailing its adaptive mechanisms—regulatory sandboxes, real-world testing, and SMEs support—and summarising their benefits in Table 1. Section 5 analyses transparency, impact assessments, accountability, and AI literacy as operational tools of responsible innovation. Section 6 engages with alternative views, addressing concerns about over-regulation and demonstrating how a well-designed framework aligns innovation with trust and legal certainty. Finally, Section 7 concludes by arguing that regulation, far from constraining progress, defines the conditions for technological ambition to endure responsibly.

\section{The False Dichotomy: Regulation vs. Innovation}


\subsection{Historic Tensions in High-Risk Sectors} 

The prevailing narrative suggests that regulation hinders innovation \cite{, Draghi2024, Ridley2020, Braben2004, Castro2019, Gikay2024, Withrow2022}, creating a false dichotomy between governance and technological progress \cite{bradford2024falsechoice, PelkmansRenda2014}. 

Historically, the United States (USA) has led AI development, buoyed by robust free markets, eminent research institutions, and a longstanding entrepreneurial ethos \cite{acemoglu2023}. Nevertheless, the new administration has revoked the Executive Order 14110 of October 30, 2023 (``Safe, Secure, and Trustworthy Development and Use of AI'') \cite{whitehouse2025}, emphasizing the removal of regulatory ``barriers'' to foster unimpeded AI innovation. This deregulatory turn is reinforced in the AI Action Plan 2025, which cautions against ``smothering'' AI in bureaucracy ``at this early stage'' and calls for ``removing red tape and onerous regulation'' to accelerate innovation \cite{AIActionPlan2025}. This approach signals a commitment to accelerate AI advancements by minimizing governmental constraints \footnote{Despite this federal deregulatory vision, several U.S. states—most notably California—have begun to adopt AI-specific legislation in recognition of the societal importance of the technology. For instance, California’s SB 53 \cite{SB53_2025} requires certain obligations for frontier models, while SB 243 \cite{SB243_2025} regulates AI companions chatbots, including disclosure and safety protocols for interactions with minors. For a state-by-state overview, see the IAPP's US State AI Governance Legislation Tracker \cite{IAPP_Tracker_2025}.}, under the premise that fewer regulations will spur competitive advantage and technological breakthroughs. However, it overlooks critical lessons from other sectors where lack of regulation has led to severe societal harms. No field with major public implications has flourished without a regulatory framework. 

In pharmaceuticals, for instance, the thalidomide scandal \cite{kingsland2020}, where over 10,000 babies were born with severe physical abnormalities due to the drug thalidomide \cite{vargesson2015}, highlighted the critical need for stringent drug regulation and clinical trial oversight. This tragedy led to the Kefauver-Harris Amendments \cite{drugamendments1962} in the USA, which mandated rigorous clinical trials and oversight for new drugs, ensuring such a disaster would not occur again. These regulations have since formed the foundation of modern pharmaceutical safety protocols. 

Additionally, the Dutch SyRI scandal \cite{rachovitsa2022, VanBekkumBorgesius2021} exemplifies the dangers of opaque AI-driven decision-making in welfare systems. In 2020, a court \cite{SyRI2020} ruled that the automated fraud detection system violated human rights due to its lack of transparency and disproportionate targeting of low-income communities, ultimately leading to the government's resignation. This case underscores the risks of unregulated AI in public administration, where biased algorithms can entrench systemic discrimination and erode trust in institutions. 

Similarly, the aviation industry once operated with far higher accident rates; in 1959, a passenger in the USA or Canada faced a 1-in-25,000 chance of a fatal crash on each departure. However, with the implementation of increasingly stringent safety regulations, aviation safety improved dramatically. The Federal Aviation Act of 1958 established the Federal Aviation Administration (FAA) \cite{faa1958}, centralizing safety oversight and enforcement in the USA. In the 1980s, the introduction of Advisory Circular AC 25.1309-1 \cite{faa13091988, faa13092024} provided systematic guidelines for aircraft system safety assessments, further reinforcing risk mitigation measures. Over time, these regulatory developments, along with international safety protocols \cite{icao1944}, contributed to an over 1,000-fold improvement in aviation safety. In the last years, the odds of a fatal accident in the USA or EU have dropped to 1 in 29 million \cite{allianz2014, icao2024, ortizospina2024}, demonstrating how robust oversight not only mitigates catastrophic risks \cite{yadav2014} but also fosters public confidence and technological innovation, as seen in modern aviation advancements. 

A failed or non‑existent regulation can stifle progress, among other things, but what these examples reflect are design flaws in the regulatory frameworks (or their absence), rather than an inherent tension between rules and advancement. Effective regulation must be thoughtfully designed to support innovation while safeguarding fundamental rights, ensuring that AI serves as a transformative force \cite{GruetzemacherWhittlestone2022} for good \cite{AI4Good2025} without compromising or even improving ethical standards. While the nature of harms differs—physical in aviation and pharmaceutics, systemic in AI—the rationale for regulation remains: to prevent irreparable consequences before they scale, ensure the security of investment, create public trust, etc. 

By doing so, a shared understanding of what constitutes valid innovation is essential; this is what the following Section is dedicated to.

\subsection{Defining Responsible Innovation} 

Traditionally, innovation has been measured by economic growth and technological novelty, often with scant attention to its social or ethical ramifications. As Smith~\cite{Smith2006} notes, the conceptual foundations of innovation stem from both management and economics, where innovation has long been seen as a driver of competitiveness and productivity. Building on Schumpeter’s~\cite{Schumpeter1934} concept of \textit{creative destruction}—the process of innovation that replaces obsolete technologies and transforms the economic structure—, formalised by Aghion and Howitt~\cite{AghionHowitt1992}, the 2025 Nobel Prize in Economics reaffirmed that sustained prosperity depends on openness to knowledge and technological renewal~\cite{NobelPrize2025}. 

 However, the landscape of innovation has evolved to incorporate broader considerations. The \textit{Oslo Manual} \cite{oecd2018} has expanded its guidelines to highlight sustainability and societal well-being as core aspects of modern innovation. This shift acknowledges that true innovation must not only advance technology and economic performance but also contribute positively to society and the environment.

In this regard, the OECD has enshrined a novel concept: responsible innovation, which means \textit{``a trustworthy technology development guided by democratic values, responsive to social needs and accountable to society. Adopting a responsible innovation approach in the development of emerging technologies can help align research and commercialisation with societal needs''} \cite{oecd2019}.

Interestingly, the revoked Executive Order 14110 \cite{whitehouseaiactions} also emphasized the promotion of responsible innovation. It's Section 2. b) stated: ``\textit{Promoting responsible innovation, competition, and collaboration will allow the USA to lead in AI and unlock the technology's potential to solve some of society's most difficult challenges}…''. Nevertheless, its revocation has signaled a shift away from these principles, potentially deprioritizing the alignment of AI  \cite{christiano2020} innovation with ethical and societal values.

In the EU context, the EU AI Act \cite{regulation2024} embodies the principles of responsible innovation by integrating regulatory oversight as a means to ensure that AI development aligns with ethical standards and societal values. The Recital 138 of the EU AI Act acknowledges that ``\textit{AI is a rapidly developing family of technologies that necessitates both a regulated environment and safe spaces for experimentation,  while ensuring responsible innovation and integration of appropriate safeguards and risk mitigation measures…}''

This responsible innovation framework upholds that rigorous governance can coexist with, and even enhance, innovation by ensuring that technological advancements do not come at the expense of societal well-being. By treating regulation as inherent to innovation, this approach demonstrates that effective governance is not a barrier but an integral part of fostering sustainable and ethical technological progress.

Consequently, in this position paper, innovation is inherently synonymous with responsible innovation, thereby rejecting the notion of innovation devoid of accountability and ethical consideration.

\subsection{Collingridge Dilemma}   

To address the issue of responsible innovation, the timely introduction of regulation plays a crucial role. The tension between early and late regulation has been characterized by Collingridge as a dilemma \cite{collingridge1980}: in the early stages of technology, it is difficult to foresee all risks, yet once the technology is deeply entrenched, change becomes exceedingly costly or even impossible. In AI, this is exacerbated by continuous deployment in high-stakes sectors and critical infrastructures where undesirable consequences may only be recognized after significant harm has occurred.

Collingridge's Dilemma highlights the challenge of regulating technology at the right time: \textit{``when change is easy, the need for it cannot be foreseen; when the need for change is apparent, change has become expensive, difficult and time consuming''} \cite{collingridge1980}. Although this captures the epistemological and societal hurdles of anticipating future harms, Collingridge does not claim the dilemma is insurmountable \cite{liebert2010}. He advocates for proactive governance mechanisms that adapt to new evidence —a perspective highly relevant to AI, where deployment in high-stakes sectors can embed problematic issues before they are fully recognized. This dilemma is not a static, universal constant; rather, it emerges from human decisions and institutional frameworks \footnote{As Carissa Véliz reminds us, ``AI is not God-given, we are designing it. We can do better.''\cite{Veliz2024}} , suggesting that deliberate regulatory strategies can and should be employed to address the inherent risks of this technology.

\section{The Risks of Deregulated AI}
\label{section:The Risks of Deregulated AI}

While some critiques frame AI regulation as a precautionary response to hypothetical risks, there is growing empirical evidence that insufficient oversight has already enabled significant and recurring harm across multiple domains \cite{incidentdatabase2025}. During the 2023 Slovak elections, deepfake audio mimicking a candidate went viral, undermining trust in the democratic process \cite{wiredSlovakia2023}. AI enables large-scale political manipulation via synthetic media, voice cloning, and micro-targeted persuasion \cite{eprs751478}.

In parallel, generative AI has enabled the spread of synthetic non-consensual sexual content, including deepfake pornography and child sexual abuse material (CSAM) \cite{unicri2024}. Notable cases include Spain’s “Almendralejo case” involving minors \cite{almendralejo2023} or AI-generated explicit images of Taylor Swift \cite{guardianSwift2024}. These cases highlight the urgent need for enforceable safeguards \cite{verfassungsblogDeepfakes2024} to prevent reputational, emotional, and psychological harm.

Beyond information harms, courts and regulators have sanctioned opaque AI systems in housing, credit, welfare, and education for violating due process and fundamental rights \cite{FPFADMReport2022}. Further risks have emerged in healthcare, financial fraud, and autonomous systems—often amplified by regulatory gaps \cite{AISafety2025}. Complementing this picture, a comparative review confirms that vulnerable groups are disproportionately affected across high-stakes domains \cite{OdiseIA2024}.

These examples show that AI-related harms are neither speculative nor isolated—they are real, ongoing, and escalating. The next sections examine two of the most urgent manifestations of these risks: bias and discrimination, and unaccountable decision-making in high-impact domains.

\subsection{Bias and Discrimination}

Recent discussions, particularly in USA, have emphasized deregulation as a means to accelerate AI development \cite{leslie2024}. However, even with existing regulations, AI presents significant risks, notably the perpetuation and amplification of societal biases \cite{mayson2019}, especially when affecting high-stakes domains. AI systems frequently learn from historical data imbued with racial \cite{cheong2024, Guardian2019DriverlessCars}, gender \cite{europeancouncil2023}, or socioeconomic biases \cite{Franklin2024}, thereby reinforcing existing disparities \cite{fra2022}. The lack of stringent regulations mandating transparent reporting and fairness audits allows these biases to remain undetected and unaddressed, potentially scaling injustices across entire populations \cite{eprs2022}. 

Biases are not purely algorithmic: human and machine decisions often reinforce each other\cite{JRC2024}, creating compounded discrimination \cite{AISafety2025}. Such dynamics highlight the complexity of mitigating discrimination, necessitating not only the identification and correction of algorithmic biases but also addressing human factors that interact with AI systems \cite{JRC2024}.

Effectively mitigating these challenges requires robust strategies bolstered by specialized tools and frameworks. A comprehensive literature review identifies 152 concrete measures for addressing and reducing biases in AI systems, encompassing interventions such as bias detection algorithms, the establishment of diverse and inclusive datasets, and rigorous auditing processes \cite{cerezo2024}. By employing these tools, organizations can systematically identify and rectify discriminatory patterns, thereby fostering fairer and more equitable AI-driven outcomes.


\subsection{Unaccountable Decision-Making in High-Risk Scenarios}

 When AI-based decisions lack regulatory oversight, affected individuals are unable to effectively challenge or seek redress for resulting harms \cite{fra2020}. The absence of humans in the loop is worrisome. But integrating human overseers without adequate training, competence or authority could be even worse, creating a false sense of legitimacy and reducing oversight to a token gesture \cite{Crootof2023,WP29_2018}. Furthermore, oversight systems frequently lack clear guidelines on when to override AI decisions, how to incorporate additional relevant information, or how to systematically audit the combined AI-human decision-making processes for cumulative biases.

This deficiency leads to ``unaccountable decision-making'', wherein neither the human operators nor the algorithms bear full responsibility for the outcomes. Such unaccountability is particularly detrimental in high-stakes scenarios, including medical diagnoses and criminal proceedings, where the concealment of biases within opaque models or data pipelines results in unjust consequences for individuals, who are then deprived of fair and transparent appeals mechanisms \cite{pasquale2015}. Effective oversight therefore requires not only technical safeguards but also structured governance mechanisms built on institutionalised distrust \cite{Laux2024} and inclusive review processes, engaging diverse stakeholders across the AI lifecycle—such as operators, auditors, end-users, and affected communities \cite{EDPS2025}.

Additionally, the phenomenon of moral outsourcing \cite{chowdhury2017, aceve2023} exacerbates this issue by transferring ethical responsibility from human decision-makers to AI systems. This occurs through the attribution of agency to AI \footnote{A scenario that could be aggravated by the emergence of AI agents. AI agents are autonomous systems capable of sensing, learning and acting upon their environments \cite{wef2024}.}, allowing both developers and deploying institutions to deflect accountability for negative outcomes onto the algorithms themselves. As a result, technical advancements are pursued as inevitable progress, while adverse effects are reframed as neutral byproducts of complex systems. This linguistic and conceptual shift places the burden of adverse outcomes on the technology itself, enabling the continued deployment of profitable AI products while perpetuating the illusion of mathematical objectivity, neutrality and accuracy \cite{oneil2017}.

Real-world examples, such as predictive policing software that utilizes racially skewed datasets \cite{larson2016compas, zuiderveenborgesius2020strengthening, buolamwini2018}, opaque credit-scoring algorithms that disadvantage low-income applicants or underrepresented minorities \cite{hurley2016}, or automated eligibility assessments in welfare and healthcare \cite{eubanks2018}, illustrate the potential for unaccountable systems and irresponsible actors to inflict irreversible harm. These systems often lack auditability, transparency, or mechanisms for individual redress, leading to decisions that are difficult to contest or revise. Once embedded into institutional infrastructures, they become resistant to scrutiny or structural change, embodying Collingridge Dilemma of being ``too late to fix''. However, regulations like the EU AI Act seek to address these risks proactively, ensuring that harm  can be mitigated before it becomes irreversible and expands.

\section{The EU AI Act: Perceived Burdens vs. Actual Benefits}
\label{subsection:The EU AI Act: Perceived Burdens vs. Actual Benefits}

\subsection{Overview of key provisions}

The proposed EU AI Act introduces a risk-based classification for AI systems \cite{Gasiola2025, ebers2024}, stipulating stricter requirements for 'high-risk' systems in fields like employment, healthcare, and education \footnote{According to the EU AI Act, this category comprises a strictly circumscribed subset of AI systems whose deployment or use may adversely affect individuals’ safety, health, or fundamental rights.}. 

 While some critics perceive these constraints as innovation-blocking \cite{Gikay2024}, the EU AI Act explicitly acknowledges the need to foster responsible AI innovation, offering mechanisms such as regulatory sandboxes and support measures for SMEs and start-ups. These tools aim to accelerate compliance through a structured and supervised environment. This section aims to debunk the myth that regulation is an impediment to innovation \cite{Tartaro2023}. On the contrary, it could become a competitive advantage.

\subsection{Regulatory sandboxes and real-world testing}

Regulatory sandboxes \cite{oecd2023a, oecd2023b, europarl2022, NovelliEtAl2025} —a concept well established in fintech \footnote{Currently, 57 countries operate 73 fintech sandboxes \cite{worldbank2020}
.}—will be mandatory in each EU Member State by August 2026, according to the Art. 57 AI Act. They provide a controlled environment where developers can train, test, and validate AI systems under regulatory supervision before the AI systems are placed on the market or otherwise put into service. Spain has already launched its first national sandbox for high-risk AI systems, offering practical insights into early implementation under real-world conditions \cite{EguiluzVia2025}.

These are not deregulated zones, but co-regulatory spaces where legal and technical innovation align to reduce uncertainty and uphold rights. This approach yields several benefits:

\begin{itemize}
    \item \textbf{Legal certainty through structured guidance}: Competent authorities provide iterative regulatory mentorship that helps participants identify and mitigate risks, clarify compliance expectations, and generate documentation that accelerates conformity assessments.
    \item\textbf{Risk mitigation}: If significant health, safety, or rights violations arise, the sandbox environment allows authorities to temporarily suspend or even end experimentation. 
    \item \textbf{No fines}: As long as companies follow the sandbox plan and act in good faith, no administrative fines apply for AI Act infringements during sandbox testing.
    \item \textbf{Cross-sectoral collaboration:} The sandbox fosters meaningful interaction between public institutions, private companies, and academia, reinforcing shared understanding and co-responsibility in the interpretation of AI obligations.
    \item \textbf{Mutual Learning}: Different actors co-construct applicable guidelines based on real evidence and case-specific interaction.

\end{itemize}

Outside these sandboxes, providers of specific high-risk AI systems—like biometrics, education, worker management—may conduct real-world testing (Art. 60 AI Act) under conditions ensuring informed consent, robust oversight, and the right to withdraw without penalty. Notably, the market surveillance authority may veto or stop tests if unaddressed risks emerge.

\subsection{Support measures for SMEs and start-ups}

Given the potentially steep costs of compliance for smaller developers, the Art. 62 of the AI Act prescribes four main forms of support:

\begin{itemize}
\item \textbf{Priority access to sandboxes}: SMEs that have either a registered office or a branch established within the EU receive preferential treatment for sandbox participation, facilitating early compliance, removing barriers and faster market. This principle is also reflected in Spain’s national sandbox, where Article 8(j) of the Royal Decree explicitly favors start-ups and SMEs in the selection process \cite{SpainSandbox2024};
\item \textbf{Awareness and training}: Member States must provide targeted outreach, such as accessible information resources and tailored training sessions on the AI Act \footnote{This includes support through initiatives like the EU’s Living Repository on AI Literacy and its accompanying FAQ guidance \cite{EULivingRepository2025,EULiteracyFAQ2025}}, to raise awareness among smaller businesses and support regulatory implementation;
\item \textbf{Dedicated communication channels}: National authorities should promptly address SMEs inquiries to minimize confusion regarding compliance procedures. Where appropriate, they may also establish dedicated communication channels to enhance clarity and support. Thus, the Commission has launched the \textit{AI Act Service Desk} and the \textit{AI Act Single Information Platform}, which include tools such as a Compliance Checker, an AI Act Explorer, and direct channels for expert guidance \cite{EUServiceDesk2025};
\item \textbf{Facilitating standardization}: SMEs should be encouraged to join standard-setting processes \cite{kilian2025, solergarrido2023, solergarrido2024}, ensuring that industry norms align with real-world developer challenges.
\end{itemize}

Furthermore, microenterprises (fewer than 10 employees, less than €2 million turnover) can implement a simplified version of the quality management system required for high-risk AI, minimizing administrative burdens without compromising user safety, as stated in the Art. 63 AI Act. Lastly, the Commission will provide standardized templates for the areas covered by the regulation and should work with Member States to lower red tape and compliance costs \cite{EUFeedback2025}.

\subsection{Competitive advantages of regulation}

Contrary to perceptions of regulatory overreach, structured regulation can bolster consumer trust and market stability \cite{bradford2024falsechoice}. 

Historical evidence from data protection (e.g., General Data Protection Regulation (GDPR) \cite{EURGDPR}) shows that companies adhering to clear legal frameworks often reap reputational benefits and avoid hefty compliance \cite{buckley2023}. While initial compliance may disrupts operations and increases costs, it ultimately enhances privacy awareness, modernizes IT infrastructure, and improves risk management processes \cite{buckley2023}. Notably, GDPR catalyzed the development of privacy-enhancing technologies (PETs), including differential privacy and federated learning \cite{OECD2025, EDPB2025}, both embedding fundamental rights into system design and reinforcing accountability and transparency \cite{bertolaccini2023, cheong2024}.


A similar dynamic is now unfolding under AI regulation. Legal obligations are accelerating the adoption of watermarking technologies to identify AI-generated content —including synthetic deepfakes— \cite{Rijsbosch2025, EPRS2023, NIST2024}. In parallel, developers are being required to implement state-of-the-art technical measures to safeguard copyright —particularly during the training of general-purpose AI models— \cite{Peukert2025} \footnote{Complying with Intellectual Property (IP) law has become one of the most pressing legal and operational challenges for the AI industry, as illustrated by the growing wave of litigation concerning training data and copyright infringement. See: Updated Map of US Copyright Suits v. AI Companies \cite{ChatGPTCopyright2025}.}. These compliance-driven innovations not only meet regulatory thresholds but also function as trust-building mechanisms that differentiate responsible actors.

Beyond technical innovation, a distinct yet complementary strategy lies in embedding ethical principles at the core of a product’s identity. A pertinent example is Apple Inc., where CEO Tim Cook strategically positioned user data privacy as both a business strategy and an ethical imperative, thereby transforming privacy into a market differentiator \cite{auyeung2021, bradford2024falsechoice}.

In light of these examples, by proactively ensuring AI meets ethical and legal standards, organizations could promote responsible innovation, while reducing the risk of reputation damage, costly litigation, and opposition from civil society. As Lucilla Sioli, head of the European Commission’s AI Office, aptly states, “\textit{You need the regulation to create trust, and that trust will stimulate innovation}'' \cite{greenacre2024}.

The strategic benefits of early compliance are well illustrated by two complementary insights: the Porter hypothesis, which holds that regulation can stimulate firms to innovate in ways that produce both societal and market benefits \cite{Porter1991, PorterVanDerLinde1995}; and the already discussed Collingridge Dilemma, which highlights the tension between limited foresight at early stages and reduced flexibility once technologies become entrenched \cite{collingridge1980}. While the dilemma does not prescribe a solution, it underscores the cost of inaction. In this context, proactive regulatory strategies—anchored in early engagement—can help mitigate long-term risks, reinforce institutional trust, and sustain the long-term viability of the AI sector, which ultimately depends on its ability to align with ethical and regulatory imperatives by design \cite{brey2024}. Firms that anticipate regulatory expectations may also capture a first-mover advantage, shape emerging markets, and convert compliance into strategic differentiation \cite{bradford2024falsechoice}.

History reinforces this logic. Industries that internalized strong safety regulations—such as medical devices or chemical manufacturing—have not only avoided crises but also secured sustained public trust and long-term leadership. Like those sectors, AI presents distinct risks when applied in sensitive domains such as healthcare, public services, and education. It must therefore follow a comparable regulatory trajectory: short-term gains from unregulated deployment may appear attractive, but risk triggering systemic failures, undermining trust, and eroding long-term viability.

Far from impeding commercial viability, the EU AI Act introduces targeted incentives—such as regulatory sandboxes, real-world testing, and SMEs support—that foster responsible experimentation under supervised conditions. These mechanisms not only facilitate compliance but also reduce deployment risks and accelerate market readiness. 

Table~\ref{tab:eu_ai_act} provides a structured comparison of common perceived burdens and their corresponding advantages under the EU AI Act. While not exhaustive, it highlights key instruments outlined above that exemplify how regulation can transform constraints into drivers of responsible AI innovation.

\begin{table}[t]
\centering
\renewcommand{\arraystretch}{1.5}
\resizebox{\textwidth}{!}{
\begin{tabular}{|c|c|}
\hline
\multicolumn{2}{|c|}{\textbf{Perceived Burdens vs. Actual Benefits of the EU AI Act}} \\
\hline
\multicolumn{2}{|c|}{\textbf{Sandboxes (Regulatory Testing \& Experimentation)}} \\
\hline 
\textit{Regulation blocks innovation} & \textbf{Regulatory Sandboxes (Art. 57)} – Safe environments for AI testing. \\
\hline
\textit{Limited flexibility for developers} & \textbf{Real-world testing (Art. 60)} – Controlled testing under oversight. \\
\hline
\multicolumn{2}{|c|}{\textbf{Support for SMEs \& Startups}} \\
\hline
\textit{High compliance costs} & \textbf{SME support (Art. 62)} – Priority access, training, and cost reduction. \\
\hline
\textit{Complex compliance for SMEs} & \textbf{Standardized templates (Art. 62)} – Simplifies compliance process. \\
\hline
\textit{Excessive bureaucracy} & \textbf{Dedicated support (Art. 62)} – Direct communication channels. \\
\hline
\multicolumn{2}{|c|}{\textbf{General Measures \& Competitive Advantages}} \\
\hline
\textit{Slows AI adoption} & \textbf{Consumer trust} – Clear rules prevent legal and market risks. \\
\hline
\textit{Limits technological progress} & \textbf{Compliance-driven innovation} – Rights-based technologies by design. \\
\hline
\textit{Stifles market growth} & \textbf{Legal certainty} – Ensures stability and reduces compliance costs. \\
\hline
\textit{Competitive disadvantage} & \textbf{Ethical leadership} – EU compliance signals responsible AI. \\
\hline
\textit{Short-term burden} & \textbf{Long-term advantage} – Reduces legal, reputational, and market risks. \\
\hline
\end{tabular}
}
\vspace{0.1em}
\caption{Perceived Burdens vs. Actual Benefits of the EU AI Act: Categorized by Key Mechanisms}
\label{tab:eu_ai_act}
\end{table}

\section{Transparency, Impact Assessments, Accountability and AI Literacy}

\label{section: Transparency, Impact Assessments, Accountability and AI Literacy}

Beyond incentives and support structures, regulation also defines how innovation is practiced. By translating legal obligations into actionable requirements and measurable design routines, governance tools such as transparency, impact assessments, accountability mechanisms, and AI literacy operationalize regulation into continuous feedback and institutional learning.

\subsection{Transparency} \label{subsection:transparency}

In a democratic context, the legitimacy of any decision-making process depends on transparency\footnote{\hspace{1mm}``\textit{With AI decision-making and its perceived legitimacy, an important question to ask is not whether we should have transparency, but rather which kind of transparency should be applied}.'' \cite{deFineLicht2020}}. The EU AI Act embodies this principle through structured documentation duties (Arts. 11, 13, and 50), ensuring that deployers, operators, and affected individuals can scrutinize AI systems, contest outcomes, and seek effective redress while aligning with fundamental rights \cite{EUFR2020AI, edpb2024, fra2022}. Building on this approach, the \textit{Code of Practice on General-Purpose AI}—endorsed by most major technology companies except Meta—extends these principles by promoting voluntary yet structured transparency commitments for general purpose AI models (Arts. 53 and 55) \cite{ec2025gpai}. Transparency—understood not merely as the provision of well-structured documentation but as the capacity to comprehend, interpret, and scrutinize the functioning of AI models—constitutes a cornerstone of trustworthy AI \cite{EUHLEG2019, Panigutti2023, Hupont2024}. 

This principle extends beyond EU borders, as reaffirmed by the Council of Europe’s Framework Convention on AI, Human Rights, Democracy and the Rule of Law \cite{coec2024}, which mandates context-sensitive transparency and oversight across the AI lifecycle. Recent proposals advance this toward meaningful transparency—information accessible and interpretable by citizens and regulators alike \cite{GaldonClavell2024}. In this sense, transparency reduces the opacity \footnote{AI models can identify complex correlations and make predictions based on vast numbers of interacting parameters, often rendering their decision-making processes opaque, even to experts. This “black box” effect can obscure the rationale behind AI-generated decisions, potentially leading to misplaced trust or over-reliance, with adverse consequences for individuals. \cite{peters2023}}, that breeds distrust, transforming compliance into a shared basis for accountability and innovation. 

\subsection{Impact Assessments}

The integration of impact assessments, such as the Fundamental Rights Impact Assessment (FRIA), represents a regulatory innovation that shifts AI governance from reactive compliance to anticipatory design. Under Article 27 of the EU AI Act, the FRIA operationalizes the protection of fundamental rights \cite{palmiotto2025} through ex ante evaluation of potential interferences. Rather than a procedural checklist, it constitutes a substantive, context-sensitive process grounded in legal and ethical expertise \cite{mantelero2024clsr, CECU2025_FRIA}.

Emerging scholarship proposes hybrid frameworks that assess the severity, duration, and reversibility of rights interferences \cite{MalgieriSantos2024}. Through supervisory oversight, the FRIA embeds proportionality, stakeholder participation, and institutional transparency, transforming abstract legal rights into measurable criteria and accountability mechanisms. Practical implementations \cite{apdcat2025, Netherlands2022_FRIA, DIHR2020_HRIA, Leslie2021_HUDERIA, AlgorithmAudit2024_FRIAReview}, show how iterative, context-aware evaluation and continuous feedback loops sustain regulatory learning—evidence that adaptive governance itself can drive responsible innovation.

\subsection{Accountability Mechanisms}

Accountability mechanisms address the structural gaps that render AI decisions unaccountable. Under Article 14 of the EU AI Act, human oversight must be exercised by competent, authorised individuals within institutional architectures that ensure traceability and reversibility. Moving beyond symbolic supervision, effective oversight relies on structured distrust—a governance principle that embeds verification and challenge throughout the AI lifecycle \cite{aepd2024, EDPB2025, Laux2024}.

Human oversight entails trained discretion, the ability to interrupt or override automated operations, and procedures for detecting and responding to adverse outcomes \cite{ec2020altai}. Yet, the deeper risk is not full automation but induced irreflexivity \cite{Innerarity2020}: automated systems press humans to decide without questioning authorship or consequence, eroding the reflective judgement. Effective governance must therefore preserve human spaces for reflexivity, to interrogate, contest and revise algorithmic outputs. When implemented through layered responsibility—linking providers, deployers, auditors and affected stakeholders—these mechanisms ensure that accountability remains both operational and reflective, sustaining trustworthy and rights-aligned innovation.

Additionally, the AI Act also foresees enforceable remedies: affected persons may file complaints with market-surveillance authorities and obtain clear, meaningful explanations for decisions that significantly impact their rights \cite{MetikosAusloos2024, Enlightenment2025_RightToExplanation}. Whistleblower protections further strengthen this ecosystem. Together, these measures ensure that accountability is not only procedural but actionable—closing the loop from oversight to redress.

\subsection{AI literacy}

AI literacy functions as the connective tissue of AI governance. Beyond technical fluency, it enables institutions and individuals to develop and use AI systems, while navigating complex regulatory landscapes and ethical dilemmas \cite{ec2024competences}. Defined in Article 3(56) of the EU AI Act, literacy encompasses the skills, knowledge, and understanding required to deploy and scrutinize AI responsibly.

Regular, practice-oriented training \cite{werneman2025}—particularly in high-risk domains—ensures that oversight remains meaningful rather than symbolic or ineffective. Through initiatives such as the EU Living Repository on AI Literacy \cite{EULivingRepository2025}, these programs transform awareness into operational capacity, allowing teams to conduct risk assessments and engage with feedback mechanisms.

Ultimately, literacy sustains the reflexive dimension of governance: it turns transparency into understanding, impact assessment into learning, and accountability into institutional memory. In this sense, AI literacy is not peripheral but constitutive of responsible innovation.

\section{Alternative Views}

Some scholars argue that regulation constrains innovation, especially in fast-moving fields like AI. The Draghi report \cite{Draghi2024} warns that excessive regulation may hinder EU digital competitiveness. Ridley \cite{Ridley2020} and Braben \cite{Braben2004} contend that scientific breakthroughs require freedom from bureaucratic and ethical constraints, and caution against premature intervention when risks are still unclear. Castro et al. \cite{Castro2019} argue that anxiety-driven regulation risks delaying adoption, constraining beneficial applications of AI, and undermining innovation itself, calling instead for optimism-driven and targeted governance frameworks. Specific critiques of the EU AI Act include Gikay’s argument that its rigid risk classification fails to balance innovation and risk \cite{Gikay2024}, and Wachter’s \cite{Wachter2024} identification of legal loopholes that may undermine enforcement and clarity.

While we acknowledge these concerns, we argue that, as documented in Section~\ref{section:The Risks of Deregulated AI}, the absence of well-designed regulation has already triggered severe harms: the viral spread of synthetic misinformation, bias and discrimination embedded in data-driven systems, and unaccountable decision-making in high-stakes contexts such as healthcare and welfare. These are not speculative threats but concrete evidence of governance failures. 

The risk-based approach adopted by the EU AI Act is therefore the appropriate regulatory model: it differentiates obligations according to risk categorization while preserving the conditions for innovation. Far from stifling innovation, it enables it through adaptive, risk-proportionate mechanisms that convert perceived burdens into competitive advantages (see table \ref{tab:eu_ai_act}). These instruments foster consumer trust, legal certainty, and ethical leadership, transforming the deregulation race into a driver of competitive trust.

The governance instruments analysed in Section~\ref{subsection:The EU AI Act: Perceived Burdens vs. Actual Benefits} and Section~\ref{section: Transparency, Impact Assessments, Accountability and AI Literacy} operationalise this balance, showing that regulation and innovation advance hand in hand. By providing the structure and accountability that make progress sustainable, the EU’s framework turns regulation into the very condition of responsible innovation.

\section{Conclusion}

This position paper has challenged the assumption that every technological advancement in AI automatically constitutes innovation. That belief ignores the extensive empirical and historical societal costs that certain AI systems have already imposed—from biased models and synthetic misinformation to unaccountable decision-making. Calling such systems “innovative” is, at best, questionable and, at worst, contradictory.

As AI becomes embedded in essential infrastructures and social decision-making, neglecting regulation in the name of ``progress'' invites the very irreversible harms that Collingridge foresaw in his Dilemma. Well-designed governance is not a brake on innovation but its catalyst—the mechanism through which technological ambition becomes sustainable, accountable, and aligned with human values. The EU AI Act exemplifies this principle: its risk-based, adaptive framework couples legal certainty with flexibility, through mechanisms such as regulatory sandboxes, small and medium enterprises (SME) support, real-world testing, fundamental rights impact assessment (FRIA), which operationalises responsible innovation.

Ultimately, innovation and regulation advance together. The future of AI will not be defined by the speed of invention but by the integrity of its governance. Regulation—when well-designed, proportionate, and rooted in fundamental rights—is the discipline that allows innovation to endure. Indeed, if AI systematically violates rights, can we truly call it innovation?

\section*{Acknowledgements}

We gratefully acknowledge Adevinta for sponsoring the industrial PhD. We also thank the Government of Catalonia’s Industrial PhDs Plan for funding part of this research. The research leading to these results has received funding from the Horizon Europe Programme under the Horizon Europe Programme under the AI4DEBUNK Project (https://www.ai4debunk.eu), grant agreement num. 101135757. Additionally, this work has been partially supported by JDC2022-050313-I  funded by MCIN/AEI/10.13039/501100011033al by European Union NextGenerationEU/PRTR.




\nocite{langley00}

\bibliography{neurips_2025}

\begin{thebibliography}{100}

\bibitem{APNews2025}
{AP News}.
\newblock {JD Vance warns that overregulation may hurt AI progress}, 2025.
\newblock Available at: \url{https://apnews.com/article/1d7826affdcdb76c580c0558af8d68d2.} Accessed October 2025.

\bibitem{BoschCEO2025}
{Times of India}.
\newblock {Bosch CEO to Europe: You are unnecessarily delaying AI’s future with overregulation}, 2025.
\newblock Available at: \url{https://timesofindia.indiatimes.com/technology/tech-news/bosch-ceo-to-europe-you-are-unnecessarily-delaying-its-ai-future-with/articleshow/122072769.cms.} Accessed October 2025.

\bibitem{MITSloan2023}
{MIT Sloan}.
\newblock {Does regulation hurt innovation? A study says yes}, 2023.
\newblock Available at: \url{https://mitsloan.mit.edu/ideas-made-to-matter/does-regulation-hurt-innovation-study-says-yes.} Accessed October 2025.

\bibitem{McAfee2021}
Andrew McAfee.
\newblock {EU Proposals to Regulate AI Are Only Going to Hinder Innovation}.
\newblock {\em Financial Times}, 2021.
\newblock Available at: \url{https://www.ft.com/content/a5970b6c-e731-45a7-b75b-721e90e32e1c.} Accessed October 2025.

\bibitem{Loten2021}
Angus Loten.
\newblock {Corporate Tech Leaders Are Mixed on EU Artificial Intelligence Bill}.
\newblock {\em Wall Street Journal}, 2021.
\newblock Available at: \url{https://www.wsj.com/articles/corporate-tech-leaders-are-mixed-on-eu-artificial-intelligence-bill-11619049736} Accessed October 2025.

\bibitem{collingridge1980}
David Collingridge.
\newblock {\em The Social Control of Technology}.
\newblock Palgrave Macmillan, 1980.

\bibitem{degregorio2022}
Giovanni De~Gregorio and Peter Dunn.
\newblock The european risk-based approaches: Connecting constitutional dots in the digital age.
\newblock {\em Common Market Law Review}, 59(2):473--500, 2022.

\bibitem{safeai2023}
{Center for AI Safety}.
\newblock Mitigating the risk of extinction from ai should be a global priority alongside other societal-scale risks such as pandemics and nuclear war, 2023.

\bibitem{RedLines2025}
{Red Lines on AI}.
\newblock Red lines on ai: A global call for ethical boundaries in artificial intelligence, 2025.
\newblock United Nations Initiative on AI Governance.

\bibitem{Draghi2024}
M.~Draghi.
\newblock Report on the future of european competitiveness, 2024.

\bibitem{Ridley2020}
M.~Ridley.
\newblock {\em How Innovation Works: And Why It Flourishes in Freedom}.
\newblock Harper, 2020.

\bibitem{Braben2004}
D.~Braben.
\newblock {\em Scientific Freedom: The Elixir of Civilization}.
\newblock Wiley, 2004.

\bibitem{Castro2019}
Daniel Castro and Michael McLaughlin.
\newblock {Ten Ways the Precautionary Principle Undermines Progress in Artificial Intelligence}, 2019.
\newblock Information Technology and Innovation Foundation. Available at: \url{https://itif.org/publications/2019/02/04/ten-ways-precautionary-principle-undermines-progress-artificial-intelligence.} Accessed October 2025.

\bibitem{Gikay2024}
Asress~Adimi Gikay.
\newblock Risks, innovation, and adaptability in the uk’s incrementalism versus the european union’s comprehensive artificial intelligence regulation.
\newblock {\em International Journal of Law and Information Technology}, 32:eaae013, 2024.

\bibitem{Withrow2022}
Josh Withrow.
\newblock Don’t stifle u.s. tech innovation with europe’s rules.
\newblock R Street Institute, October 9 2022.

\bibitem{bradford2024falsechoice}
Anu Bradford.
\newblock The false choice between digital regulation and innovation.
\newblock {\em Northwestern University Law Review}, 118(2), October 6 2024.
\newblock {Available at: https://dx.doi.org/10.2139/ssrn.4753107.} Accessed October 2025.

\bibitem{PelkmansRenda2014}
Jacques Pelkmans and Andrea Renda.
\newblock Does eu regulation hinder or stimulate innovation?
\newblock CEPS Special Report~96, Centre for European Policy Studies (CEPS), November 2014.
\newblock Available at SSRN: https://ssrn.com/abstract=2528409.

\bibitem{acemoglu2023}
Daron Acemoğlu and Simon Johnson.
\newblock {\em Power and Progress: Our Thousand-Year Struggle Over Technology and Prosperity}.
\newblock Basic Books, 2023.

\bibitem{whitehouse2025}
President\_of\_USA.
\newblock Removing barriers to american leadership in artificial intelligence.
\newblock Executive Order, January 23 2025.

\bibitem{AIActionPlan2025}
The~White House.
\newblock America's ai action plan, 2025.
\newblock Office of Science and Technology Policy, Executive Office of the President.

\bibitem{SB53_2025}
California Legislature.
\newblock Senate bill no. 53, artificial intelligence models: Large developers.
\newblock \url{https://leginfo.legislature.ca.gov/faces/billTextClient.xhtml?bill_id=202520260SB53}, 2025.
\newblock Chaptered on September 29, 2025.

\bibitem{SB243_2025}
California Legislature.
\newblock Senate bill no. 243, companion chatbots.
\newblock \url{https://leginfo.legislature.ca.gov/faces/billNavClient.xhtml?bill_id=202520260SB243}, 2025.
\newblock Chaptered on October 13, 2025.

\bibitem{IAPP_Tracker_2025}
International~Association of~Privacy~Professionals.
\newblock U.s. state ai governance legislation tracker, 2025.
\newblock Accessed October 2025.

\bibitem{kingsland2020}
James Kingsland.
\newblock How the thalidomide scandal led to safer drugs.
\newblock {\em Medical News Today}, 2020.

\bibitem{vargesson2015}
Neil Vargesson.
\newblock Thalidomide-induced teratogenesis: History and mechanisms.
\newblock {\em Journal of Developmental Biology}, 2015.

\bibitem{drugamendments1962}
{United States Congress}.
\newblock Public law 87-781 - drug amendments of 1962, October 10 1962.

\bibitem{rachovitsa2022}
Aristi Rachovitsa and Nathalie Johann.
\newblock The human rights implications of the use of ai in the digital welfare state: Lessons learned from the dutch syri case.
\newblock {\em Human Rights Law Review}, 22(2), 2022.

\bibitem{VanBekkumBorgesius2021}
Marvin van Bekkum and Frederik Zuiderveen~Borgesius.
\newblock Digital welfare fraud detection and the dutch syri judgment.
\newblock {\em European Journal of Law and Technology}, 23(4), 2021.
\newblock First published online August 2, 2021.

\bibitem{SyRI2020}
{NJCM et al. v The Dutch State}.
\newblock The hague district court ecli:nl:rbdha:2020:1878 (syri).
\newblock The Hague District Court, 2020.
\newblock English translation available at: \url{http://deeplink.rechtspraak.nl/uitspraak?id=ECLI:NL:RBDHA:2020:1878.} Accesed October 2025.

\bibitem{faa1958}
{Federal Aviation Administration}.
\newblock The federal aviation act of 1958, 1958.

\bibitem{faa13091988}
{Federal Aviation Administration}.
\newblock Advisory circular 25.1309-1a - system design and analysis, 1988.

\bibitem{faa13092024}
{Federal Aviation Administration}.
\newblock Advisory circular 25.1309-1b - system design and analysis, 2024.

\bibitem{icao1944}
{International Civil Aviation Organization (ICAO)}.
\newblock Convention on international civil aviation, December 7, 1944.

\bibitem{allianz2014}
Allianz.
\newblock Global aviation safety study, 2014.

\bibitem{icao2024}
{International Civil Aviation Organization (ICAO)}.
\newblock Safety report, 2024.

\bibitem{ortizospina2024}
Esteban Ortiz-Ospina.
\newblock Commercial flights have become significantly safer in recent decades.
\newblock {\em Our World in Data}, 2024.

\bibitem{yadav2014}
D.K. Yadav and H.~Nikraz.
\newblock Implications of evolving civil aviation safety regulations on the safety outcomes of air transport industry and airports.
\newblock {\em Aviation}, 18:94--103, 2014.

\bibitem{GruetzemacherWhittlestone2022}
Ross Gruetzemacher and Jess Whittlestone.
\newblock The transformative potential of artificial intelligence.
\newblock {\em Futures}, 135, 2022.

\bibitem{AI4Good2025}
{AI for Good Foundation}.
\newblock Ai for good, 2025.
\newblock Available at: https://ai4good.org/. Accessed October 2025.

\bibitem{Smith2006}
Keith Smith.
\newblock Measuring innovation.
\newblock In Jan Fagerberg, David~C. Mowery, and Richard~R. Nelson, editors, {\em The Oxford Handbook of Innovation}, pages 148--178. Oxford University Press, Oxford, 2006.

\bibitem{Schumpeter1934}
Joseph~A. Schumpeter.
\newblock {\em The Theory of Economic Development: An Inquiry into Profits, Capital, Credit, Interest, and the Business Cycle}.
\newblock Harvard University Press, Cambridge, MA, 1934.

\bibitem{AghionHowitt1992}
Philippe Aghion and Peter Howitt.
\newblock A model of growth through creative destruction.
\newblock {\em Econometrica}, 60(2):323--351, 1992.

\bibitem{NobelPrize2025}
{Royal Swedish Academy of Sciences}.
\newblock Scientific background: The sveriges riksbank prize in economic sciences in memory of alfred nobel 2025.
\newblock \url{https://www.nobelprize.org/prizes/economic-sciences/2025/press-release/}.
\newblock Accessed October 2025.

\bibitem{oecd2018}
{OECD/Eurostat}.
\newblock {\em Oslo Manual 2018: Guidelines for Collecting, Reporting and Using Data on Innovation, 4th Edition}.
\newblock OECD Publishing, Paris/Eurostat, Luxembourg, 2018.

\bibitem{oecd2019}
{OECD Science, Technology and Industry Working Papers}.
\newblock Responsible innovation in neurotechnology enterprises.
\newblock October 11 2019.

\bibitem{whitehouseaiactions}
{Biden-Harris Administration}.
\newblock Fact sheet: Biden-harris administration announces key ai actions following president biden's landmark executive order, 2024.
\newblock {Available at: www.whitehouse.gov/briefing-room/statements-releases/2024/01/29/fact-sheet-biden-harris-administration-announces-key-ai-actions-following-president-bidens-landmark-executive-order/.} Accesed October 2025.

\bibitem{christiano2020}
{Brian Christian}.
\newblock {\em The Alignment Problem: Machine Learning and Human Values}.
\newblock W. W. Norton \& Company, 2020.

\bibitem{regulation2024}
{European Parliament and Council}.
\newblock Regulation (eu) 2024/1689 on artificial intelligence (artificial intelligence act).
\newblock June 13 2024.

\bibitem{liebert2010}
Wolfgang Liebert and Jan~C. Schmidt.
\newblock Collingridge’s dilemma and technoscience.
\newblock {\em Poiesis and Praxis}, 7(1-2):55--71, 2010.

\bibitem{Veliz2024}
Carissa Véliz.
\newblock How privacy can save your life.
\newblock YouTube.
\newblock {Available at: www.youtube.com/watch?v=xSPRouBvgFE.} Accesed October 2025.

\bibitem{incidentdatabase2025}
{Partnership on AI}.
\newblock Ai incident database.
\newblock \url{https://incidentdatabase.ai/}, 2025.
\newblock Accessed October 2025.

\bibitem{wiredSlovakia2023}
Morgan Meaker.
\newblock Slovakia’s election deepfakes show ai is a danger to democracy.
\newblock \url{https://www.wired.com/story/slovakias-election-deepfakes-show-ai-is-a-danger-to-democracy/}, 2023.
\newblock Accessed October 2025.

\bibitem{eprs751478}
{European Parliamentary Research Service}.
\newblock Artificial intelligence, democracy and elections.
\newblock Technical Report EPRS\_BRI(2023)751478, European Parliament, 2023.
\newblock Briefing Paper.

\bibitem{unicri2024}
{Bracket Foundation and Value for Good GmbH in cooperation with the Centre for Artifi cial Intelligence and Robotics at the United Nations Interregional Crime and Justice Research Institute (UNICRI).}
\newblock Generative ai: A new threat for online child sexual exploitation and abuse.
\newblock Technical report, UNICRI, 2024.
\newblock Accessed October 2025.

\bibitem{almendralejo2023}
{BBC News}.
\newblock Ai-generated naked child images shock spanish town of almendralejo.
\newblock \url{https://www.bbc.com/news/world-europe-66877718}, 2023.
\newblock Accessed October 2025.

\bibitem{guardianSwift2024}
Emine Saner.
\newblock Inside the taylor swift deepfake scandal: 'it’s men telling a powerful woman to get back in her box'.
\newblock \url{https://www.theguardian.com/technology/2024/jan/31/inside-the-taylor-swift-deepfake-scandal-its-men-telling-a-powerful-woman-to-get-back-in-her-box}, 2024.
\newblock Accessed October 2025.

\bibitem{verfassungsblogDeepfakes2024}
Beatriz Kira.
\newblock Deepfakes, the weaponisation of ai against women and possible solutions, 2024.
\newblock Published on Verfassungsblog, availabe at: https://verfassungsblog.de/deepfakes-ncid-ai-regulation/. Accessed October 2025.

\bibitem{FPFADMReport2022}
Sebastião~Barros Vale and Gabriela Zanfir-Fortuna.
\newblock Automated decision-making under the gdpr: Practical cases from courts and data protection authorities.
\newblock Report, Future of Privacy Forum, May 2022.
\newblock Accessed on January 31, 2024.

\bibitem{AISafety2025}
UK~Government and International~Expert Panel.
\newblock International ai safety report 2025.
\newblock {Available at: https://www.gov.uk/government/publications/international-ai-safety-report-2025/international-ai-safety-report-2025.} Accessed October 2025.

\bibitem{OdiseIA2024}
OdiseIA.
\newblock Inventory of comparative judgments and rulings: Ai and vulnerable groups, 2024.

\bibitem{leslie2024}
David Leslie and Andrea~M. Perini.
\newblock Future shock: Generative ai and the international ai policy and governance crisis.
\newblock {\em Harvard Data Science Review}, Special Issue 5, 2024.

\bibitem{mayson2019}
Sandra Mayson.
\newblock Bias in, bias out.
\newblock {\em Yale Law Journal}, 128:2218--2300, 2019.

\bibitem{cheong2024}
B.C. Cheong.
\newblock Transparency and accountability in ai systems: Safeguarding wellbeing in the age of algorithmic decision-making.
\newblock {\em Frontiers in Human Dynamics}, 6:1421273, 2024.

\bibitem{Guardian2019DriverlessCars}
The Guardian.
\newblock The racism of technology - and why driverless cars could be the most dangerous example yet.
\newblock The Guardian (Shortcuts blog), March 13 2019.

\bibitem{europeancouncil2023}
{European Council}.
\newblock Study on the impact of artificial intelligence systems, their potential for promoting equality, including gender equality, and the risks they may cause in relation to non-discrimination, 2023.

\bibitem{Franklin2024}
G.~Franklin, R.~Stephens, M.~Piracha, S.~Tiosano, F.~Lehouillier, R.~Koppel, and P.~L. Elkin.
\newblock The sociodemographic biases in machine learning algorithms: A biomedical informatics perspective.
\newblock {\em Life (Basel)}, 14(6):652, May 21 2024.

\bibitem{fra2022}
{European Union Agency for Fundamental Rights (FRA)}.
\newblock Bias in algorithms - artificial intelligence and discrimination, December 8 2022.

\bibitem{eprs2022}
{European Parliamentary Research Service (EPRS)}.
\newblock Auditing the quality of datasets used in algorithmic decision-making systems, 2022.

\bibitem{JRC2024}
A.~Gaudeul, O.~Arrigoni, V.~Charisi, M.~Escobar~Planas, and I.~Hupont~Torres.
\newblock The impact of human oversight on discrimination in ai-supported decision-making: A large case study on human oversight of ai-based decision support systems in lending and hiring scenarios, 2024.

\bibitem{cerezo2024}
Pablo Cerezo-Martínez, Alejandro Nicolás-Sánchez, and Francisco~J. Castro-Toledo.
\newblock Analyzing the european institutional response to ethical and regulatory challenges of artificial intelligence in addressing discriminatory bias.
\newblock {\em Frontiers in Artificial Intelligence}, 7, June 25 2024.

\bibitem{fra2020}
{European Union Agency for Fundamental Rights (FRA)}.
\newblock Getting the future right – artificial intelligence and fundamental rights, December 14 2020.

\bibitem{Crootof2023}
Rebecca Crootof, Margot~E. Kaminski, and W.~Nicholson Price~II.
\newblock Humans in the loop.
\newblock {\em Vanderbilt Law Review}, 76:429--494, 2023.
\newblock Also available as U of Colorado Law Legal Studies Research Paper No. 22-10 and U of Michigan Public Law Research Paper No. 22-011.

\bibitem{WP29_2018}
{Article 29 Data Protection Working Party}.
\newblock Guidelines on automated individual decision-making and profiling for the purposes of regulation 2016/679 (wp251rev.01), 2018.
\newblock Adopted on 3 October 2017, as last revised and adopted on 6 February 2018.

\bibitem{pasquale2015}
Frank Pasquale.
\newblock {\em The Black Box Society: The Secret Algorithms That Control Money and Information}.
\newblock Harvard University Press, 2015.

\bibitem{Laux2024}
Jens Laux.
\newblock Institutionalised distrust and human oversight of artificial intelligence: Towards a democratic design of ai governance under the european union ai act.
\newblock {\em AI \& Society}, 39:2853--2866, 2024.

\bibitem{EDPS2025}
{European Data Protection Supervisor}.
\newblock Techdispatch: Human oversight of automated decision-making.
\newblock Technical report, European Data Protection Supervisor (EDPS), 2025.
\newblock Issue Authors: Vítor Bernardo, Laura Hernández. EDPS Technology and Privacy Unit.

\bibitem{chowdhury2017}
Rumman Chowdhury.
\newblock Moral outsourcing: Finding the humanity in artificial intelligence.
\newblock {\em Forbes}, 2017.

\bibitem{aceve2023}
Paula Aceve.
\newblock ``i do not think ethical surveillance can exist''': Rumman chowdhury on accountability in ai.
\newblock {\em The Guardian}, 2023.

\bibitem{wef2024}
World~Economic Forum.
\newblock Navigating the ai frontier: A primer on the evolution and impact of ai agents, 2024.

\bibitem{oneil2017}
Cathy O’Neil.
\newblock {\em Weapons Of Math Destruction: How Big Data Increases Inequality and Threatens Democracy}.
\newblock 2017.

\bibitem{larson2016compas}
Jeff Larson, Julia Angwin, Lauren Kirchner, and Surya Mattu.
\newblock How we analyzed the compas recidivism algorithm.
\newblock {\em ProPublica}, May 2016.
\newblock Accessed October 2025.

\bibitem{zuiderveenborgesius2020strengthening}
F.~J. Zuiderveen~Borgesius.
\newblock Strengthening legal protection against discrimination by algorithms and artificial intelligence.
\newblock {\em The International Journal of Human Rights}, 24(10):1572--1593, 2020.

\bibitem{buolamwini2018}
Joy Buolamwini and Timnit Gebru.
\newblock Gender shades: Intersectional accuracy disparities in commercial gender classification.
\newblock {\em Proceedings of Machine Learning Research}, 81:77--91, 2018.

\bibitem{hurley2016}
Michelle Hurley and Julius Adebayo.
\newblock Credit scoring in the era of big data.
\newblock {\em Yale Journal of Law and Technology}, 18(1):148, 2016.

\bibitem{eubanks2018}
Virginia Eubanks.
\newblock {\em Automating Inequality: How High-Tech Tools Profile, Police, and Punish the Poor}.
\newblock St. Martin’s Press, 2018.

\bibitem{Gasiola2025}
Gustavo~Gil Gasiola.
\newblock Rebuilding the pyramid: The ai act’s risk-based approach using a binary decision diagram.
\newblock {\em Computer Law \& Security Review}, 58:106189, 2025.

\bibitem{ebers2024}
Martin Ebers.
\newblock Truly risk-based regulation of artificial intelligence: How to implement the eu’s ai act.
\newblock {\em European Journal of Risk Regulation}, 2024.

\bibitem{Tartaro2023}
A.~Tartaro, A.~L. Smith, and P.~Shaw.
\newblock Assessing the impact of regulations and standards on innovation in the field of ai.
\newblock {\em arXiv preprint arXiv:2302.04110}, 2023.

\bibitem{oecd2023a}
OECD.
\newblock Regulatory sandboxes in artificial intelligence, July 13 2023.

\bibitem{oecd2023b}
OECD.
\newblock Regulatory sandboxes can facilitate experimentation in artificial intelligence, May 31 2023.

\bibitem{europarl2022}
European Parliament.
\newblock Artificial intelligence act and regulatory sandboxes, June 2022.

\bibitem{NovelliEtAl2025}
Claudio Novelli, Philipp Hacker, Simon McDougall, Jessica Morley, Antonino Rotolo, and Luciano Floridi.
\newblock Getting regulatory sandboxes right: Design and governance under the ai act, 2025.
\newblock Available at SSRN.

\bibitem{worldbank2020}
The World~Bank Group.
\newblock Global experiences from regulatory sandboxes, 2020.

\bibitem{EguiluzVia2025}
Josu~A. Eguiluz~Castañeira and Anna Via.
\newblock First european ai high-risk sandbox lessons: How smes can implement high-risk ai in a real-world environment, 2025.
\newblock SSRN Working Paper.

\bibitem{SpainSandbox2024}
Ministry of~Economic~Affairs and Digital Transformation.
\newblock Royal decree 817/2023 of 8 november establishing a controlled testing environment for the trial of compliance with the proposal for a regulation of the european parliament and of the council laying down harmonised rules on artificial intelligence, 2024.
\newblock Published in the Official State Gazette (BOE) No. 268, 9 November 2023, Article 8(j).

\bibitem{EULivingRepository2025}
European Commission.
\newblock Living repository to foster learning and exchange on ai literacy, 2025.
\newblock Accessed October 2025.

\bibitem{EULiteracyFAQ2025}
European Commission.
\newblock Ai literacy – questions and answers, 2025.
\newblock Accessed October 2025.

\bibitem{EUServiceDesk2025}
European Commission.
\newblock Commission launches ai act service desk and single information platform to support ai act, 2025.
\newblock Accessed October 2025.

\bibitem{kilian2025}
Robert Kilian, Linda Jäck, and Dominik Ebel.
\newblock European ai standards – technical standardisation and implementation challenges under the eu ai act.
\newblock {\em European Journal of Risk Regulation}, pages 1--25, 2025.

\bibitem{solergarrido2023}
J.~Soler Garrido, D.~Fano Yela, C.~Panigutti, H.~Junklewitz, R.~Hamon, T.~Evas, A.~André, and S.~Scalzo.
\newblock Analysis of the preliminary ai standardisation work plan in support of the ai act, 2023.

\bibitem{solergarrido2024}
J.~Soler~Garrido, S.~De~Nigris, E.~Bassani, I.~Sanchez, T.~Evas, A.~André, and T.~Boulangé.
\newblock Harmonised standards for the european ai act.
\newblock Technical Report JRC139430, European Commission, Seville, 2024.

\bibitem{EUFeedback2025}
European Commission.
\newblock Commission collects feedback to simplify rules on data, cybersecurity and artificial intelligence in upcoming initiatives, 2025.
\newblock Accessed October 2025.

\bibitem{EURGDPR}
European Parliament.
\newblock Regulation on the protection of natural persons with regard to the processing of personal data and on the free movement of such data, and repealing directive 95/46/ec (general data protection regulation), December 2016.

\bibitem{buckley2023}
Gerard Buckley, Tristan Caulfield, and Ingolf Becker.
\newblock “it may be a pain in the backside but…” insights into the resilience of business after gdpr.
\newblock {\em Proceedings of the 2022 New Security Paradigms Workshop (NSPW '22)}, pages 21--34, 2023.

\bibitem{OECD2025}
Organisation for Economic Co-operation and Development.
\newblock Privacy-enhancing technologies (pets), 2025.
\newblock Accessed October 2025.

\bibitem{EDPB2025}
European Data~Protection Board.
\newblock Training curriculum on ai and data protection fundamentals of secure ai systems with personal data, 2025.
\newblock See Section 5: Privacy Enhancing Technologies in AI Systems.

\bibitem{bertolaccini2023}
L.~Bertolaccini, P.E. Falcoz, A.~Brunelli, J.~Furák H.F.~Batirel, S.~Passani, and Z.~Szanto.
\newblock The significance of general data protection regulation in the compliant data contribution to the european society of thoracic surgeons database.
\newblock {\em European Journal of Cardio-Thoracic Surgery}, 2023.

\bibitem{Rijsbosch2025}
Konrad~Kollnig Bram~Rijsbosch, Gijs van~Dijck.
\newblock Adoption of watermarking measures for ai-generated content and implications under the eu ai act.
\newblock arXiv preprint, 2025.

\bibitem{EPRS2023}
European Parliamentary~Research Service.
\newblock Generative ai and watermarking.
\newblock Briefing Document, European Parliament, 2023.
\newblock Accessed October 2025.

\bibitem{NIST2024}
National~Institute of~Standards and Technology (NIST).
\newblock Reducing risks posed by synthetic content: An overview of technical approaches to digital content transparency, 2024.
\newblock NIST AI 100-4, Trustworthy and Responsible AI Series.

\bibitem{Peukert2025}
Alexander Peukert and Céline Castets-Renard.
\newblock Code of practice for general-purpose ai models: Copyright chapter.
\newblock European Commission, 2025.
\newblock Working Group 1, Copyright Chapter, includes obligations under Article 53(1)(c) AI Act.

\bibitem{ChatGPTCopyright2025}
{ChatGPT is Eating the World}.
\newblock Updated map of us copyright suits v. ai companies (oct. 19, 2025), 2025.
\newblock Accessed October 2025.

\bibitem{auyeung2021}
Angela Au-Yeung.
\newblock Apple forces competitors to play by its rules with a new operating system.
\newblock {\em Forbes}, April 20 2021.

\bibitem{greenacre2024}
Martin Greenacre.
\newblock Eu is ‘losing the narrative battle’ over ai act, says un adviser.
\newblock {\em Science|Business}, December 5 2024.

\bibitem{Porter1991}
Michael~E. Porter.
\newblock America's green strategy.
\newblock {\em Scientific American}, pages 168--176, April 1991.

\bibitem{PorterVanDerLinde1995}
Michael~E. Porter and Claas van~der Linde.
\newblock Toward a new conception of the environment–competitiveness relationship.
\newblock {\em Journal of Economic Perspectives}, 9(4):97--118, 1995.

\bibitem{brey2024}
Philip Brey and Benjamin Dainow.
\newblock Ethics by design for artificial intelligence.
\newblock {\em AI Ethics}, 4:1265–1277, 2024.

\bibitem{deFineLicht2020}
Katrin de~Fine~Licht and Johannes de~Fine~Licht.
\newblock Artificial intelligence, transparency, and public decision-making.
\newblock {\em AI \& Society}, 35:917--926, 2020.

\bibitem{EUFR2020AI}
{European Union Agency for Fundamental Rights}.
\newblock Getting the future right – artificial intelligence and fundamental rights.
\newblock Technical report, European Union Agency for Fundamental Rights, December 2020.

\bibitem{edpb2024}
{European Data Protection Board (EDPB)}.
\newblock Edpb opinion on ai models: Gdpr principles support responsible ai, December 18 2024.

\bibitem{ec2025gpai}
{European Commission}.
\newblock The general-purpose ai code of practice, July 2025.
\newblock Accessed: 2025-10-22.

\bibitem{EUHLEG2019}
{High-Level Expert Group on Artificial Intelligence}.
\newblock Ethics guidelines for trustworthy ai.
\newblock European Commission, April 2019.

\bibitem{Panigutti2023}
Cecilia Panigutti, Ronan Hamon, Isabelle Hupont, David~Fernandez Llorca, Delia~Fano Yela, Henrik Junklewitz, Salvatore Scalzo, Gabriele Mazzini, Ignacio Sanchez, Josep~Soler Garrido, and Emilia Gomez.
\newblock The role of explainable ai in the context of the ai act.
\newblock In {\em Proceedings of the 2023 ACM Conference on Fairness, Accountability, and Transparency (FAccT '23)}, pages 1--12, New York, NY, USA, 2023. ACM.

\bibitem{Hupont2024}
Isabelle Hupont, David Fernandez-Llorca, Silvia Baldassarri, et~al.
\newblock Use case cards: A use case reporting framework inspired by the european ai act.
\newblock {\em Ethics and Information Technology}, 26:19, 2024.

\bibitem{coec2024}
Council\_of\_Europe.
\newblock Framework convention on artificial intelligence and human rights, democracy and the rule of law, September 5 2024.

\bibitem{GaldonClavell2024}
Gemma~Galdon Clavell.
\newblock Ai auditing: Proposal for ai leaflets.
\newblock European Data Protection Board, SPE Programme, 2024.
\newblock Accessed October 2025.

\bibitem{peters2023}
U.~Peters.
\newblock Explainable ai lacks regulative reasons: Why ai and human decision-making are not equally opaque.
\newblock {\em AI and Ethics}, 3(3):963--974, 2023.

\bibitem{palmiotto2025}
Francesco Palmiotto.
\newblock The ai act roller coaster: The evolution of fundamental rights protection in the legislative process and the future of the regulation.
\newblock {\em European Journal of Risk Regulation}, pages 1--24, 2025.

\bibitem{mantelero2024clsr}
Alessandro Mantelero.
\newblock The fundamental rights impact assessment (fria) in the ai act: Roots, legal obligations and key elements for a model template.
\newblock {\em Computer Law \& Security Review}, 54:106020, 2024.

\bibitem{CECU2025_FRIA}
Confederación~Española de~Consumidores~y Usuarios~(CECU).
\newblock Towards a meaningful implementation of article 27 under the ai act: Participation in fundamental rights impact assessments (frias) \& effective transparency mechanisms for their oversight and enforcement.
\newblock Technical report, CECU, 2025.
\newblock Accessed October 2025.

\bibitem{MalgieriSantos2024}
Gianclaudio Malgieri and Cristiana Santos.
\newblock Assessing the (severity of) impacts on fundamental rights.
\newblock {\em Computer Law \& Security Review}, 2025.
\newblock Forthcoming.

\bibitem{apdcat2025}
{Catalan Data Protection Authority}.
\newblock Fria model: Guide and use cases. fria methodology for ai design and development, 2025.

\bibitem{Netherlands2022_FRIA}
{Government of the Netherlands}.
\newblock Impact assessment fundamental rights and algorithms.
\newblock Technical report, Ministry of the Interior and Kingdom Relations, 2021.
\newblock Accessed October 2025.

\bibitem{DIHR2020_HRIA}
{Danish Institute for Human Rights}.
\newblock Guidance on human rights impact assessment of digital activities.
\newblock Technical report, Danish Institute for Human Rights, 2020.
\newblock Accessed October 2025.

\bibitem{Leslie2021_HUDERIA}
David Leslie, Christopher Burr, et~al.
\newblock Human rights, democracy, and the rule of law assurance framework for ai systems: A proposal (huderia).
\newblock Technical report, The Alan Turing Institute, 2022.
\newblock Accessed October 2025.

\bibitem{AlgorithmAudit2024_FRIAReview}
{Algorithm Audit}.
\newblock A comparative review of 10 fundamental rights impact assessments (fria) for ai systems.
\newblock Technical report, Algorithm Audit, 2024.
\newblock Accessed October 2025.

\bibitem{aepd2024}
{Spanish Data Protection Agency (AEPD)}.
\newblock Evaluating human intervention in automated decisions, 2024.

\bibitem{ec2020altai}
{Independent High-Level Expert Group on Artificial Intelligence}.
\newblock The assessment list for trustworthy artificial intelligence (altai) for self-assessment, 2020.

\bibitem{Innerarity2020}
Daniel Innerarity.
\newblock El impacto de la inteligencia artificial en la democracia.
\newblock {\em Revista de las Cortes Generales}, (109):87--103, 2020.

\bibitem{MetikosAusloos2024}
Ljubiša Metikoš and Jef Ausloos.
\newblock The right to an explanation in practice: Insights from case law for the gdpr and the ai act.
\newblock {\em Law, Innovation and Technology}, pages 1--36, March 2025.

\bibitem{Enlightenment2025_RightToExplanation}
{Research Institute – Digital Human Rights Center}.
\newblock The right to explanation in the ai act business manual.
\newblock Technical report, Research Institute – Digital Human Rights Center, Vienna, Austria, 2025.
\newblock Project: Enlightenment 4.0 – Human Understanding of AI Decision-Making.

\bibitem{ec2024competences}
R.~Medaglia, P.~Mikalef, and L.~Tangi.
\newblock Competences and governance practices for artificial intelligence in the public sector, 2024.

\bibitem{werneman2025}
E.~Werneman Root and M.~Mahay.
\newblock Assessing ai literacy needs.
\newblock {\em IAPP}, 2025.

\bibitem{Wachter2024}
Sandra Wachter.
\newblock Limitations and loopholes in the eu ai act and ai liability directives: What this means for the european union, the united states, and beyond.
\newblock {\em Yale Journal of Law and Technology}, 26(3), 2024.

\bibitem{langley00}
P.~Langley.
\newblock Crafting papers on machine learning.
\newblock In Pat Langley, editor, {\em Proceedings of the 17th International Conference on Machine Learning (ICML 2000)}, pages 1207--1216, Stanford, CA, 2000. Morgan Kaufmann.

\end{thebibliography}
\bibliographystyle{unsrt}

\end{document}